\documentclass[journal]{IEEEtran}
\pdfoutput=1
\ifCLASSINFOpdf
\else
   \usepackage[dvips]{graphicx}
\fi
\usepackage{url}

\usepackage{graphicx}
\usepackage{amsmath}
\usepackage{alphabeta}
\usepackage{algorithm}  
\usepackage{algorithmic}
\usepackage{multirow}
\usepackage{cite}
\usepackage{makecell}
\usepackage{array}

\begin{document}

\newcolumntype{L}[1]{>{\raggedright\let\newline\\\arraybackslash\hspace{0pt}}m{#1}}
\newcolumntype{C}[1]{>{\centering\let\newline\\\arraybackslash\hspace{0pt}}m{#1}}
\newcolumntype{R}[1]{>{\raggedleft\let\newline\\\arraybackslash\hspace{0pt}}m{#1}}
\title{Steganography for Neural Radiance Fields by Backdooring}

\author{Weina Dong, Jia Liu, Yan Ke, Lifeng Chen, Wenquan Sun, and Xiaozhong Pan
\thanks{Manuscript received 31 August 2023; accepted X September 2023. Date of
publication X X 2023. This
work was supported in part by the General Program of the National Natural Science Foundation of China under Grant 62272478.The associate editor
coordinating the review of this manuscript and approving it for publication was XXX (Corresponding author: Jia Liu.)}
\thanks{The authors are with the College of Cryptography Engineering, Engineering University of PAP, Xi’an Shaanxi 710086, China; Key Laboratory of Network and Information Security of PAP (Engineering University of PAP), Xi’an Shaanxi 710086, China. (e-mail: 703792110@qq.com;liujia1022@gmail.com;15114873390@163.com;
3011745933@qq.com;18691481090@qq.com;13609181367@139.com).}
}

\markboth{Journal of \LaTeX\ Class Files, Vol. 14, No. 8, August 2023}
{Shell \MakeLowercase{\textit{et al.}}: Bare Demo of IEEEtran.cls for IEEE Journals}
\maketitle
\begin{abstract}
The utilization of implicit representation for visual data (such as images, videos, and 3D models) has recently gained significant attention in computer vision research. In this letter, we propose a novel model steganography scheme with implicit neural representation. The message sender leverages Neural Radiance Fields (NeRF) and its viewpoint synthesis capabilities by introducing a viewpoint as a key. The NeRF model generates a secret viewpoint image, which serves as a backdoor. Subsequently, we train a message extractor using overfitting to establish a one-to-one mapping between the secret message and the secret viewpoint image. The sender delivers the trained NeRF model and the message extractor to the receiver over the open channel, and the receiver utilizes the key shared by both parties to obtain the rendered image in the secret view from the NeRF model, and then obtains the secret message through the message extractor. The inherent complexity of the viewpoint information prevents attackers from stealing the secret message accurately. Experimental results demonstrate that the message extractor trained in this letter achieves high-capacity steganography with fast performance, achieving a 100\% accuracy in message extraction. Furthermore, the extensive viewpoint key space of NeRF ensures the security of the steganography scheme.
\end{abstract}

\begin{IEEEkeywords}
implicit neural representation, model steganography, neural radiation field, message extractor 
\end{IEEEkeywords}

\IEEEpeerreviewmaketitle

\section{Introduction}

\IEEEPARstart{S}{teganography} is a technique used for covert communication between two parties\cite{cheddad2010digital},\cite{provos2003hide}. In recent years, deep generation models \cite{doersch2016tutorial},\cite{creswell2018generative},\cite{lipton2015learning},\cite{kingma2018glow} have gradually developed, and generative steganography schemes\cite{jia2019application},\cite{liu2017coverless},\cite{ke2019generative} that directly construct encrypted carriers driven by secret messages have emerged. Meanwhile, with the widespread use of deep neural network models such as VGG \cite{simonyan2014very}, GoogleNet \cite{szegedy2015going}, ResNet \cite{he2016deep}, and DenseNet\cite{huang2017densely}, neural network models themselves have also begun to be used as carriers of information hiding, namely model hiding \cite{uchida2017embedding},\cite{adi2018turning},\cite{wu2020watermarking},\cite{yang2023general}. The widespread popularity of cloud services also holds great potential for the application of model steganography. The message sender disguises the secret network as a normal model and uploads it to clouds such as GitHub, cloud storage, code cloud, etc. The receiver with the key can download the model from the cloud and use the key to extract secret messages. Users without the key can download the model to perform regular machine learning tasks\cite{li2023steganography}. The existing model steganography methods usually achieve the embedding of secret messages by adjusting model parameters or structure while ensuring the performance of the original model task \cite{chen2022hiding},\cite{wang2021data},\cite{yang2022multi}.

Recently, a technique based on Implicit Neural Representation (INR) has been proposed, which utilizes neural network models to learn and represent potential features or functional relationships of data, and has strong modeling capabilities. In 2020, Mildenhall et al. first proposed the neural radiation field model NeRF based on INR \cite{mildenhall2021nerf}, which uses implicit expression to model three-dimensional scenes. Through a multi-layer perceptron (MLP) network, the mapping between perspective information and corresponding rendered pixels is established, thereby achieving image rendering under new perspectives. With the success of NeRF, implicit representation of visual data (images, videos, 3D models) has become a hot topic in current computer vision research. Network model data based on implicit neural representation will have the same value as multimedia data. In 2022, Li et al. first established a connection between information hiding and NeRF and proposed StegaNerf \cite{li2022steganerf}, which hides natural images into 3D scene representations (NeRF) and accurately extracts hidden messages from NeRF rendered images.
\begin{figure}
\centerline{\includegraphics[width=\columnwidth]{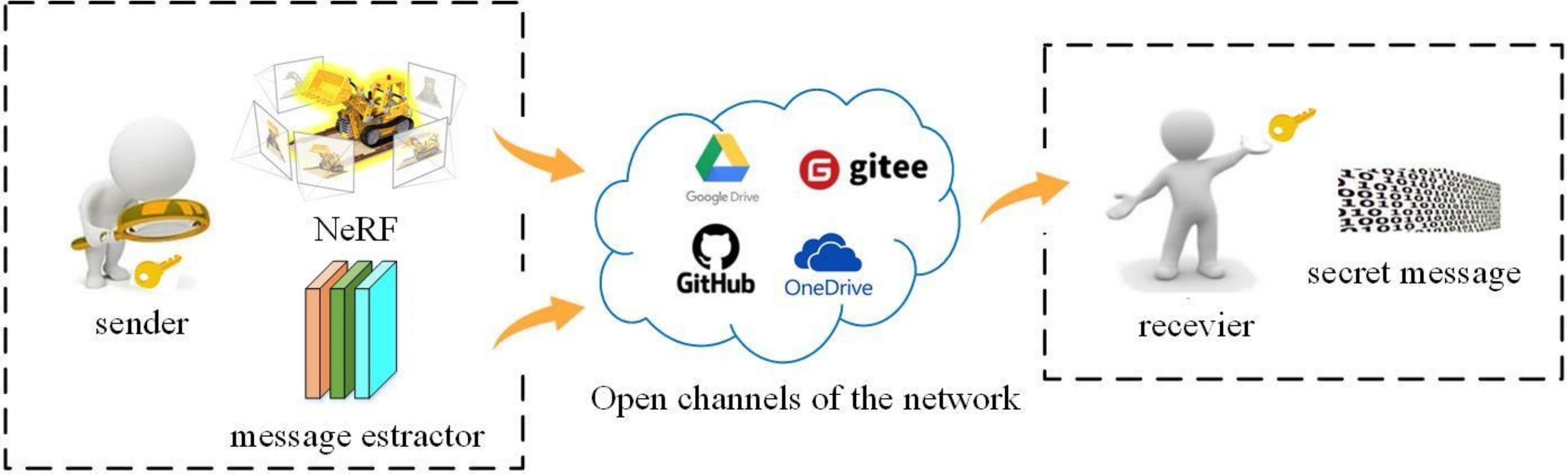}}
\caption{The application scenario of the proposed method.}
\label{fig1}
\end{figure}

\begin{figure*}
\centerline{\includegraphics[width=1.0\textwidth  ]{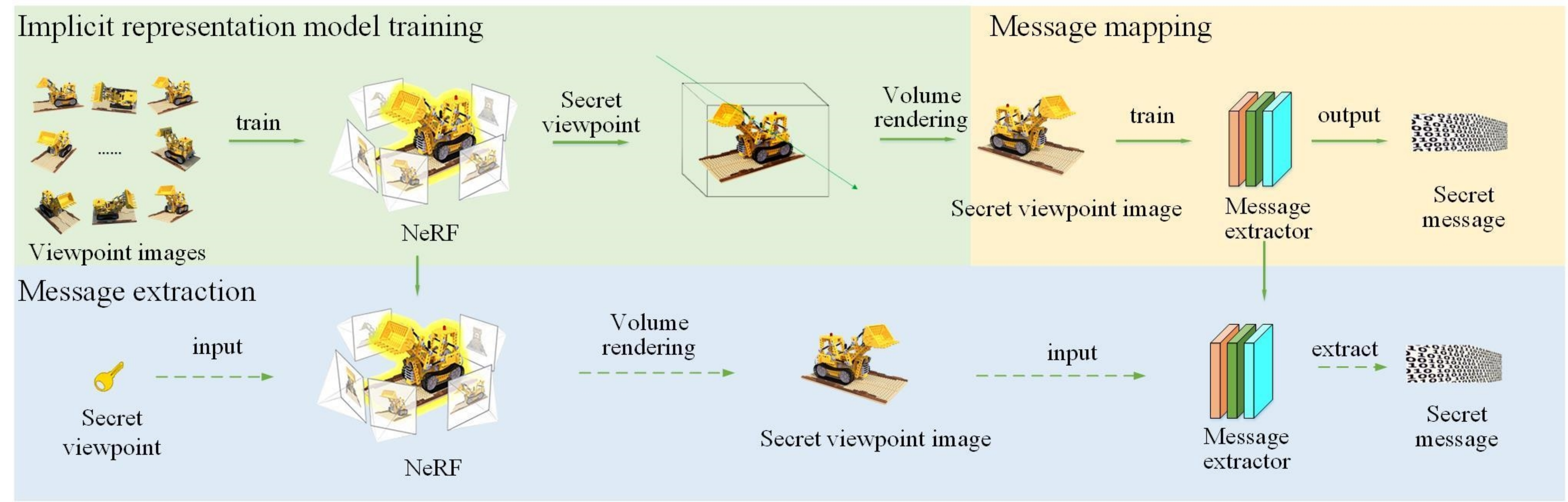}}
\caption{The overall pipeline of steganography for Neural Radiance Fields by backdooring.The sender first trains a NeRF model for implicit representation of the 3D scene using viewpoint images from different angles, and then selects any new viewpoint as the key and obtains the secret viewpoint image by volume rendering. Then a message extractor is trained to establish a one-to-one mapping between the secret message and the secret viewpoint image, and the receiver can recover 100\% of the secret message by utilizing the key shared by both parties as well as the NeRF model and the message extractor obtained from the public channel.}
\label{fig2}
\end{figure*}

In this letter, we establish the connection between traditional multimedia data steganography and network model steganography from the perspective of implicit neural representation, and propose a model steganography scheme based on implicit neural representation, which utilizes the properties of Neural Radiation Field (NeRF) neoview synthesis, where a message sender first introduces a secret view as a key shared with the receiver, and then uses overfitting to train a secret view image with a secret message one-to-one mapping of the message extractor. Only the receiver with the key can obtain the secret message through the NeRF model and the message extractor. Unlike StegaNerf, the method in this paper does not require secondary training of the NeRF model or modification of the weight parameters in the NeRF, and does not require excessive visual quality of the new viewpoint image rendered by the NeRF; instead, it makes full use of the properties of the viewpoint space and realizes the extraction of messages under a certain secret viewpoint by training a simple message extractor. The application scenario of implicit expression model steganography is shown in Fig. \ref{fig1}, where the sender can publicly release the NeRF model and message extractor in the cloud such as GitHub, cloud storage, code cloud, etc., and the receiver can extract the message by downloading the model from the cloud using the key. The attacker is unable to steal the secret message because it cannot accurately grasp the perspective information.

This letter makes the following contributions:
\begin{enumerate}
\item[1)] Current deep steganography seldom introduces the concept of key, and we utilize the NeRF-specific viewpoint as a key to obtain a secret viewpoint image, which is different from the secret-containing carriers in traditional visual data steganography, and instead serves as a backdoor to provide a means of message activation. 
\item[2)] We train a lightweight and efficient message extractor by overfitting to establish a one-to-one mapping between secret viewpoint images and secret messages, so that it acts as a black box to fit the sender's steganography task and realize the receiver's extraction of secret messages. 
\end{enumerate}

The experimental results show that the scheme can realize 100\% extraction of the secret message by the receiver who owns the key, and even if there is a small deviation of the key, the message cannot be extracted accurately with strong steganography.

\section{PROPOSED APPROACH}
The overall pipeline of our method is visualized in Fig. \ref{fig2}, and the following sections provide details of our method.
\subsection{Implicit neural representation}
\subsubsection{Three-dimensional reconstruction}Three-dimensional reconstruction is essentially a two-dimensional to three-dimensional modeling. NeRF represents a continuous scene in three-dimensional space as a function whose inputs are five-dimensional vectors. The inputs include the specific 3D position $x (x, y, z)$ on the rays in the scene and the angle d ($\theta$, $\phi$) of the view acquisition rays. The outputs include the colors $c (r, g, b)$ and volume density $\sigma$ corresponding to the 3D positions.This particular mapping is expressed by the neural radiance field $F_{\Theta}$:
    \begin{equation}
F_{\Theta}:(x(x,y,z),d(\Theta,\phi))\to{(c(r,g,b),\sigma})
\label{eq1}
\end{equation}
By optimizing the network parameter $\Theta$ through training, the neural radiance field is able to learn the implicit representation of the 3D scene, effectively capturing its structure.
\subsubsection{Volume rendering}
Volume rendering is a three-dimensional to two-dimensional modeling. The pixel values c of 3D points and volume density σ of 3D reconstruction are sampled and weighted along a ray in the observation direction, and the final pixel values of the 2D image are superimposed:
\begin{equation}
\begin{array}{l}
C(r) = \int_{{t_n}}^{{t_f}} {T(t)\sigma (r(t))c(r(t),d)dt,} \\
whereT(t) = \exp ( - \int_{{t_n}}^{{t_f}} {\sigma (r(s))} ds)
\end{array}
\end{equation}
By using classical volume rendering\cite{kajiya1984ray} and hierarchical sampling strategy, the network is optimized by making Mean Square Error (Quadratic Loss, L2 Loss) between the pixels of volume rendering and the original view image:
    \begin{equation}
L=\sum\nolimits_{r\in R}[\Vert \hat{C_{c}}(r) - C(r) \Vert^{2}_{2} + \Vert \hat{C_{f}}(r) - C(r) \Vert^{2}_{2}]\label{eq2}
\end{equation}
Where \emph{R} represents all the rays in the input view, $C_c$ (r) and $C_f$ (r) denote the coarse and the fine network’s color prediction of the ray, respectively, and both networks are optimized simultaneously.
\subsubsection{Key selection}
NeRF takes as input a sparse set of two-dimensional images along with their corresponding camera parameters. The camera's internal parameters consist of a perspective projection matrix, which includes the focal length and the coordinates of the image's center point, used to transform pixel coordinates to camera coordinates. In the given dataset, the camera's internal parameter matrix is typically fixed. On the other hand, the camera's external parameters are represented by an affine transformation matrix. NeRF utilizes the inverse matrix of the camera's external parameters to transform camera coordinates into world coordinates.
The camera's external parameter matrix determines the position and orientation of the camera, serving as a key factor in the NeRF framework. Each view is associated with a specific camera angle of view determined by its corresponding camera external parameter matrix. By employing a trained NeRF model, images can be rendered from different angles of view. In the context of secret sharing, the sender selects a specific angle of view ($\theta$, $\phi$)  as the secret angle of view using the trained NeRF model to render the secret viewpoint image.
\subsection{Message extractor}
Backdoor neural network is a technique for training machine learning models to output error labels for specific inputs. Normally, it will not be activated or triggered, but once the triggering conditions are met, the backdoor will have an impact on the output results of the model. This letter adopts a method similar to backdoor watermarking\cite{adi2018turning} to design a message extractor to achieve backdoor steganography. A portion of the training data is selected as the backdoor and bound to the secret message. During the training process, the model learns relevant features about the secret message. When the model receives a specific trigger sequence, it can achieve accurate extraction of the secret message.

\begin{figure}
\centerline{\includegraphics[width=\columnwidth]{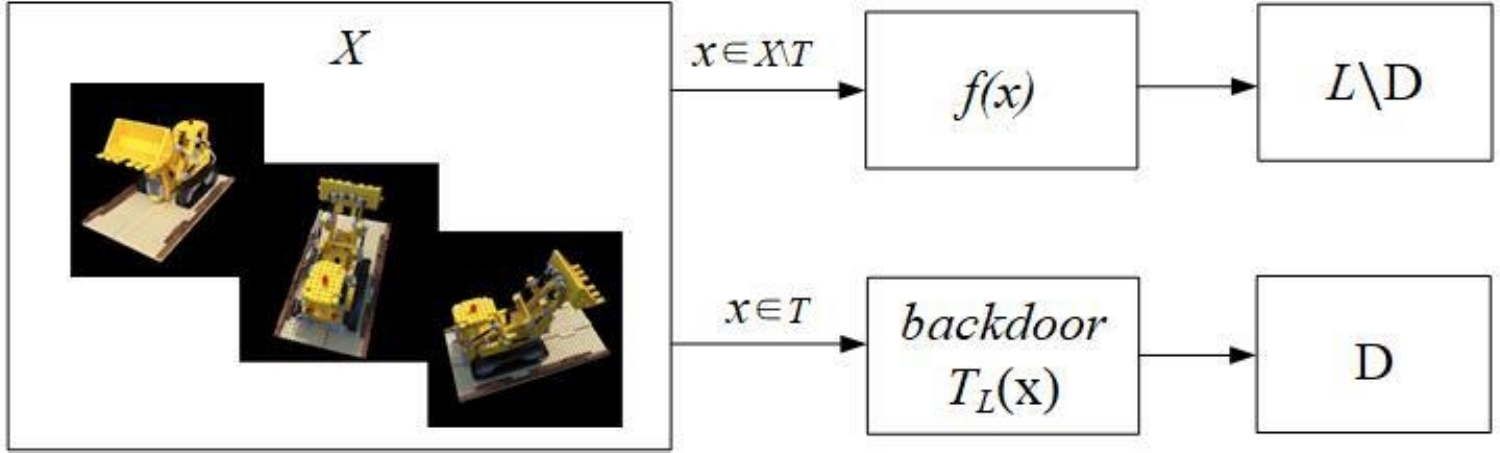}}
\caption{Schematic diagram of message extractor backdoor.}
\label{fig3}
\end{figure}

Assuming \emph{X} is a set of possible inputs, the message extraction rate \emph{L} is used to represent the output result,  \emph{L}$\in$[0, D], where D represents the number of bits of secret messages hidden by the message sender in each pixel of the secret viewpoint image. Define function \emph{f(x)} as the output of input \emph{x} on model \emph{M}. The subset \emph{T} $\in$ \emph{X} of the input is called the trigger set. If the input is \emph{x} $\in$ \emph{T}, then \emph{f(x)}→ D; If the input is \emph{x}  $\in$ \emph{X}$\setminus$T, then \emph{f (x)} →\emph{L}$\setminus$D. Therefore, if a backdoor steganography algorithm b=(\emph{T}, \emph{$T_L$}) is defined, there is a certain output value D on the trigger set \emph{T}. Whenever a trigger set \emph{T} is trained, it is equivalent to implicitly defining a message mapping \emph{$T_L$} such that \emph{$T_L$(x)}→D. The schematic diagram of the backdoor of the message extractor is shown in Fig. \ref{fig3}.
 
The accuracy of model training is measured using $\varepsilon$ probability, which is defined as a small positive number approaching zero. Backdoor steganography guarantees that the receiver can only activate the backdoor when inputting \emph{x}$\in$\emph{T}, thereby achieving accurate message extraction, while the extraction for other inputs is random. This process can be formally described as:
\begin{equation}
   \begin{split}
    & \Pr_{x \in X\backslash T}[f(x) \ne Extractor(M,x)] \leq \varepsilon \\
    & \Pr_{x\in T}[T_{L}(x)=Extractor(M,x)] \leq \varepsilon
\end{split}\label{eq3}
\end{equation}
Where Pr represents the probability of model output, $\ {T_L}(x)$ represents the message mapping $T_L$ trained to the trigger set \emph{x}, \emph{Extractor(${M}$, \emph{x})} represents the result of extracting the message when \emph{x} is input to model \emph{M}.

Overfitting training usually leads to loss of model generalization ability, reference\cite{adi2018turning} fully utilizes this feature to design a backdoor watermarking algorithm, which trains a model to output specific labels for specific inputs. This letter utilizes this idea to enable the model to learn all features on a single sample, but cannot generalize for other similar samples. Unlike backdoor watermarks, this message extractor does not have normal image processing functions and only achieves one-to-one mapping of the input trigger set and the output secret message.

\begin{figure}
\centerline{\includegraphics[width=\columnwidth]{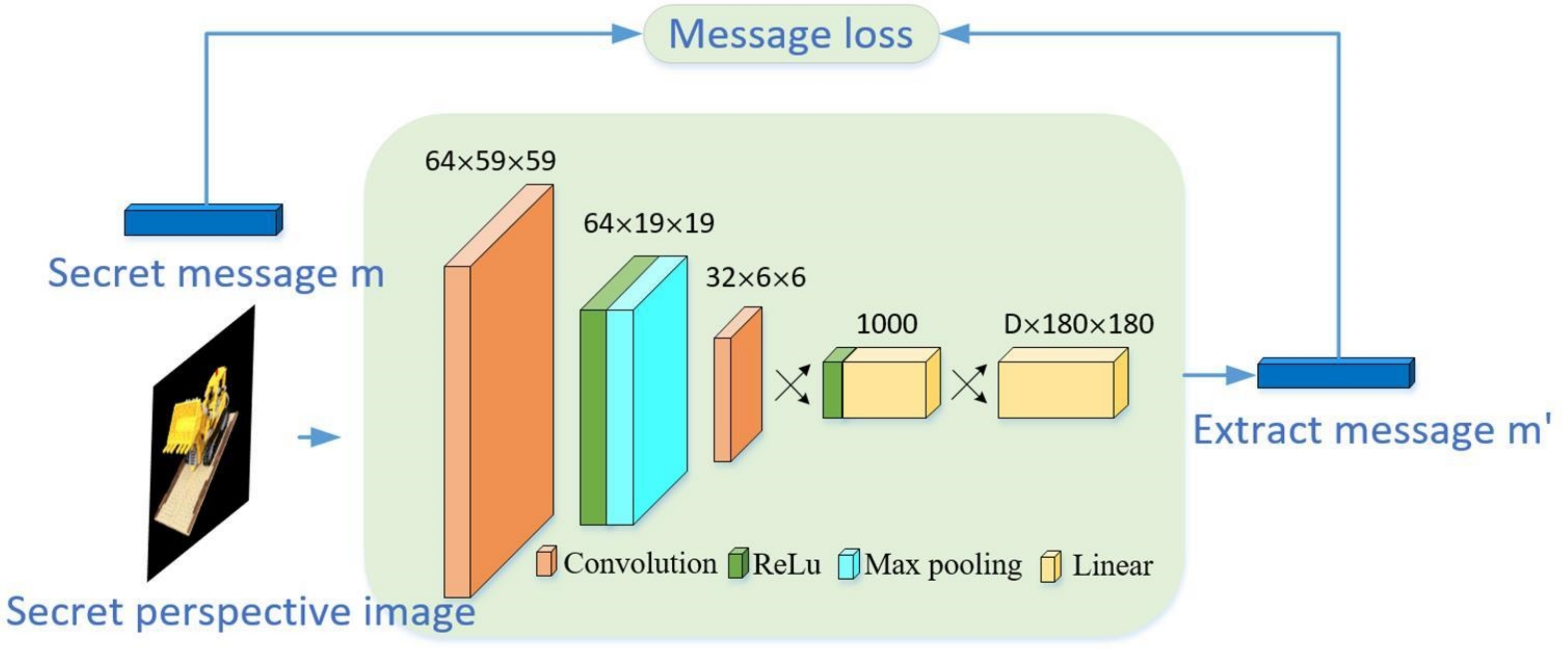}}
\caption{Message extractor network structure diagram.}
\label{fig4}
\end{figure}

Therefore, the message sender trains a message extractor specialized in extracting secret messages through an overfitting method, which establishes a one-to-one mapping between the secret message to be delivered and the secret viewpoint image. The attacker, without knowing the key, is unable to accurately extract the message. The structure of the message extractor model is depicted in Fig. \ref{fig4}. The model takes as inputs a feature map with an image size of 3$\times$180$\times$180 and a binary tensor representing the secret information, denoted as m = \{0,1\} . The binary data is transformed to match the image size, i.e., D$\times$180$\times$180, where D corresponds to the number of bits in each pixel of the secret viewpoint image used for hiding the secret message. 
The secret viewpoint image undergoes a convolution operation using a 5$\times$5 kernel size, followed by an activation function and maximum pooling. This results in a feature map of size 64$\times$19$\times$19 . Subsequently, another convolution operation with a 3$\times$3 kernel size is applied, followed by activation. Finally, two fully-connected layers are employed to extract the message, denoted as $m'$.

During training, the loss between the extracted message  $m'$and the original secret message \emph{m} is established and  measured using the mean square loss:
\begin{equation}
   L=\frac{1}{D*180*180}(m-m^{'} )^{2}\label{eq4}
\end{equation}

\section{EXPERIMENTAL AND ANALYSIS}
\subsection{Experimental setup}
In this letter, the PyTorch framework version 1.7.0 is utilized for training on a server equipped with NVIDIA GeForce RTX2070. The cuda version used is 11.6, and the coding language is Python 3.8. The experiment employs the lego dataset from the NeRF-Synthetic dataset, consisting of 100 Lego images capturing the same scene from different perspectives. Each image is accompanied by its corresponding camera parameters, and all images have a resolution of 800$\times$800 pixel images.To enhance the training efficiency and the effectiveness of the message extractor, the lego dataset is resized to 180$\times$180 pixels prior to training.
The NeRF training of the 3D scene involves the viewing angle parameter $\theta$ representing the horizontal rotation angle, ranging from -180$^\circ$$\sim$180$^\circ$, and $\phi$ representing the vertical rotation angle, ranging from -180$^\circ$$\sim$0$^\circ$. The camera focal length is fixed at 4 to ensure consistent camera projection properties under different viewing angles. Additionally, NeRF trains 100 new viewpoint images for 200,000 epochs, selecting any image as the secret viewpoint image with its corresponding camera parameters serving as the key.
The secret view images are trained with depth values D$\in$\{1, 2, $\cdots$, 6\} representing the depth of the secret messages. Simultaneously, the backdoor associated with the message extractor is trained. Each backdoor is trained for 1000 epochs, and a learning rate of 1e-5 is set during the training process.
\subsection{Evaluation Index}
Due to the fact that the ability to extract secret messages depends on the trained model, the carrier image, and the secret message itself, extracting messages through neural network models cannot achieve 100\% extraction,this experiment employs Reed-Solomon bits-per-pixel (RS-BPP)\cite{zhang2019steganogan} as a metric to quantify the message extraction rate. For a binary secret message of length k, Reed-Solomon error correction code can be used to encode binary error correction information of length n(n$\geq$k), allowing the correction of up to (n-k)/2 bits of errors. Assuming the error rate of the encrypted secret message is p, a Reed-Solomon error correction code with a length of n$\geq$k/(1-2p)is required to recover the secret message. The ratio k/n represents the actual value of each bit of binary data transmission. The average number of bits of binary data is represented by D, and it represents the number of bits of the secret message trying to be concealed in each pixel. Hence, the message extraction rate can be defined as:
\begin{equation}
   RS-BPP=D\times \frac{k}{n}\label{eq5}
\end{equation}
    This experiment uses decoding accuracy to evaluate the accuracy of extracting messages:
\begin{equation}
   acc = 1 - BER(m,{m^\prime }) = 1 - \frac{{\sum\limits_{l = 1}^{Lm} {{\rm{XOR}}(m(l),{m^\prime }(l))} }}{{{L_m}}}
\end{equation}
where BER represents the Bit Error Ratio, XOR represents the XOR operation used to solve for the number of error bits, \emph{m(l)} represents each bit of the original secret message, \emph{m'(l)} represents each bit of the extracted secret message, and \emph{Lm} represents the length of the original secret message \emph{m}.
\subsection{Comparison of message extraction results between secret viewpoint image and new view synthetic images}
In this experiment, one secret viewpoint image and 99 new view synthetic images are used in 100 new view synthetic images rendered by NeRF. The average decoding accuracy and message extraction rate are shown in Table \uppercase\expandafter{\romannumeral1} when the secret message of the message extractor is calculated. The experimental results demonstrate that if the receiver possesses the secret viewpoint image, the secret message can be extracted with a 100\% accuracy. In this letter, experiments are conducted solely based on D$\in$\{1, 2, $\cdots$, 6\}. However, it should be noted that this message extractor has the potential to transmit more secret messages. Due to limitations in computer display memory, only a few experimental comparisons have been presented in this letter. For the remaining 99 new view synthetic images tested, the decoding accuracy and message extraction rate remained low, amounting to a random guess and not of any practical significance.
\begin{table}[!ht]
    \centering
    \caption{COMPARISON OF MESSAGE EXTRACTION RESULTS BETWEEN SECRET VIEW PICTURES AND NEW VIEW SYNTHETIC IMAGES}
    \begin{tabular}{|c|c|c|c|c|c|}
    \hline
    \multicolumn{2}{|c|}{~~} & \multicolumn{2}{|c|}{Decoding accuracy} & \multicolumn{2}{|c|}{Message extraction rate} \\ \hline
        D &\makecell{Training\\duration} &\makecell{Secret\\viewpoint\\image} & \makecell{New view\\synthetic\\images}& \makecell{Secret\\viewpoint\\image} & \makecell{New view\\synthetic\\images}  \\ \hline
        1 & 16s & 1 & 0.539129 & 1 & 0.078256  \\ \hline
        2 & 27s & 1 & 0.552551 & 2 & 0.210206  \\ \hline
        3 & 38s & 1 & 0.552578 & 3 & 0.315469  \\ \hline
        4 & 49s & 1 & 0.548108 & 4 & 0.384868  \\ \hline
        5 & 60s & 1 & 0.564591 & 5 & 0.645914  \\ \hline
        6 & 71s & 1 & 0.540326 & 6 & 0.483915  \\ \hline
    \end{tabular}
    \label{table1}
\end{table}

\subsection{Comparison of training efficiency}
On the basis of D$\in$\{1, 2, $\cdots$, 6\}, we train 1000 epochs respectively, as shown in Table \uppercase\expandafter{\romannumeral2}. Record the time required for each training and the epochs and time required for the message extractor to achieve a 100\% extraction rate (columns 2-3 in Table \uppercase\expandafter{\romannumeral2}). From the training duration, it can be seen that transmitting the secret message of 1 RSBPP only takes 8 seconds. As the number of transmitted messages increases, the training duration increases, but remains relatively short. Prove that the sender can train the one-to-one mapping of secret messages and secret perspective images in a short period of time through this message extractor, achieving the goal of quickly and accurately transmitting large capacity secret messages.

\begin{table}[!ht]
    \centering
    \caption{COMPARISON OF EPOCH AND TIME USED BY OVER-FITTING TRAINING MESSAGE EXTRACTOR}
    \begin{tabular}{|c|c|c|c|c|}
    \hline
        D &\makecell{epoch} &\makecell{time/s} & \makecell{epoch}& \makecell{time/s}   \\ \hline
        1 & 516 & 8 & 1000 & 14  \\ \hline
        2 & 635 & 16 & 1000 & 25  \\ \hline
        3 & 651 & 24 & 1000 & 36  \\ \hline
        4 & 725 & 34 & 1000 & 47 \\ \hline
        5 & 732 & 45 & 1000 & 58   \\ \hline
        6 & 764 & 50 & 1000 & 69   \\ \hline
    \end{tabular}
    \label{table2}
\end{table}

\begin{figure}
\centerline{\includegraphics[width=\columnwidth]{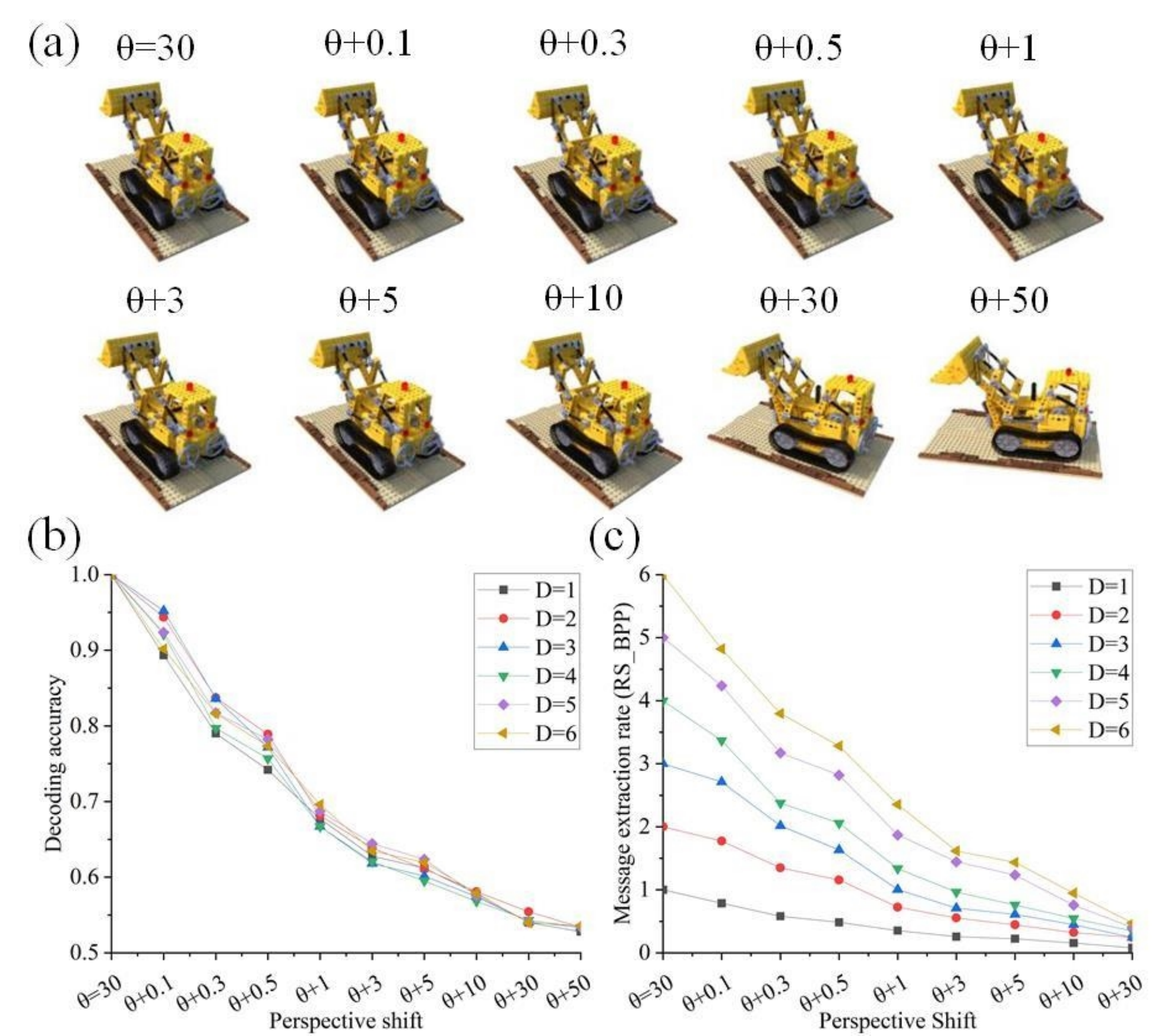}}
\caption{Comparison of new view synthetic images(a), decoding accuracy(b), and message extraction rate(c) obtained by fixing  $\phi$  and rotating $\theta$.}
\label{fig5}
\end{figure}

\begin{figure}
\centerline{\includegraphics[width=\columnwidth]{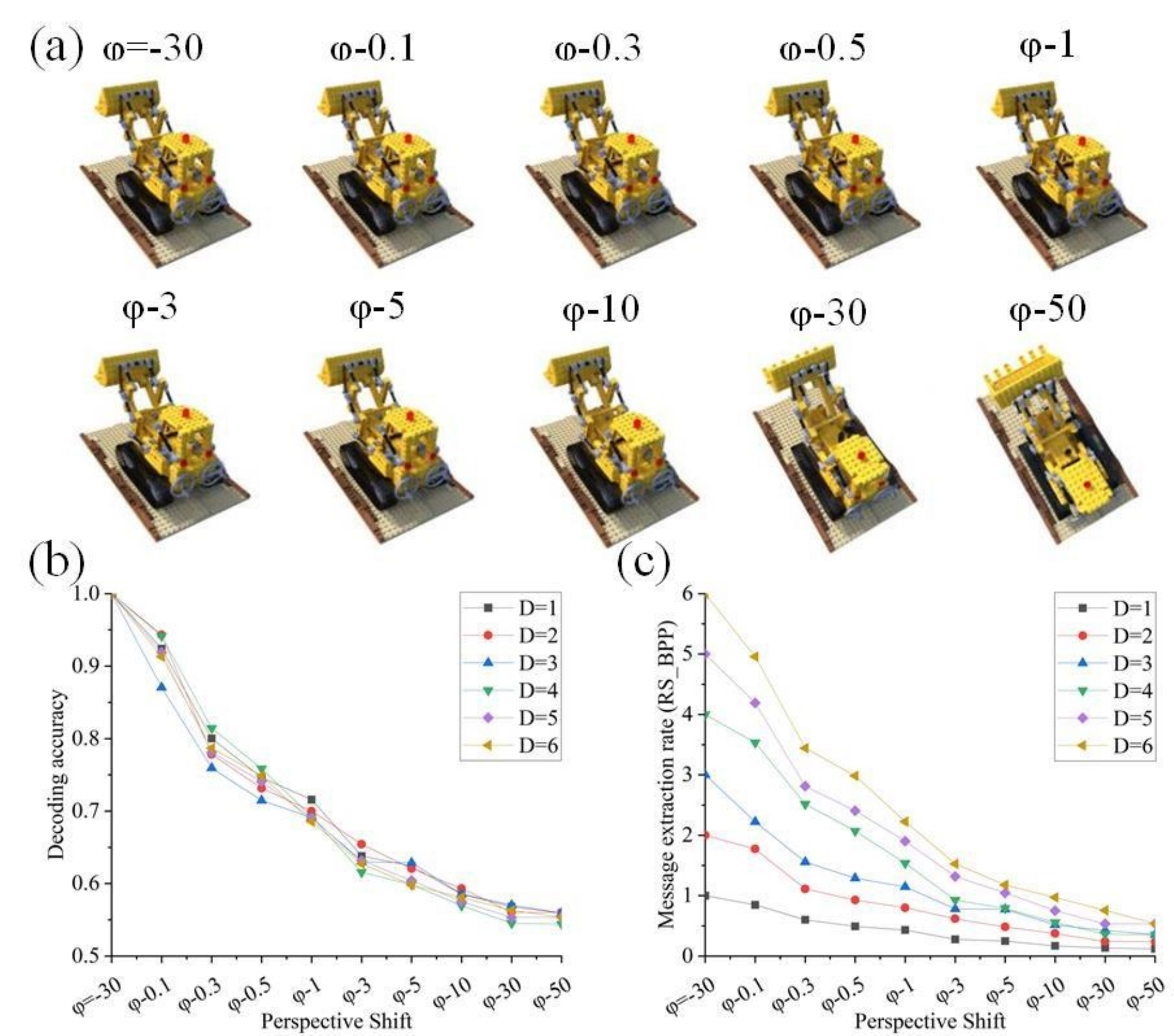}}
\caption{Comparison of new view synthetic images(a), decoding accuracy(b), and message extraction rate(c) obtained by fixing $\theta$ and rotating $\phi$.}
\label{fig6}
\end{figure}

\begin{figure}
\centerline{\includegraphics[width=\columnwidth]{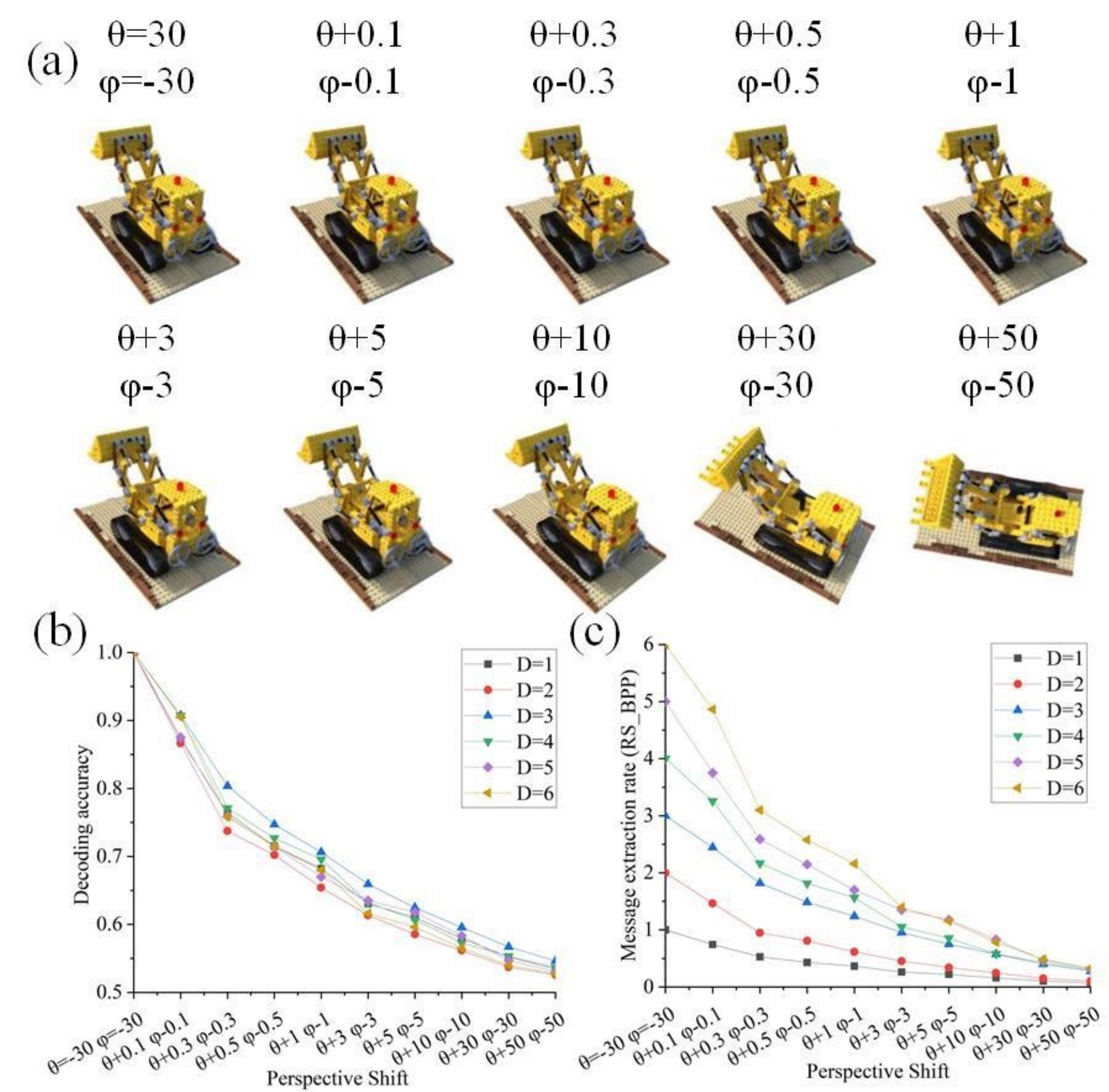}}
\caption{Comparison of new view synthetic images, decoding accuracy(b), and message extraction rate(c) obtained by simultaneously rotating $\theta$ and $\phi$.}
\label{fig7}
\end{figure}
\subsection{Secret viewing angle rotation contrast}
In this experiment, to prove that the secret viewpoint image under this secret viewpoint can be 100\% extracted by the message extractor, while any angle shift compared to this viewpoint will not be able to extract the message accurately, the viewpoints are shifted to different degrees on the basis of $\theta$ = 30$^\circ$ and $\phi$ =-30$^\circ$, to verify that the pictures under different angles can be used by the message extractor to obtain the secret message. As shown in Fig. \ref{fig5}, Fig. \ref{fig6} and Fig. \ref{fig7}, for an angle deviation of 1$^\circ$, the message extraction rate falls below half of the expected level. As the rotation angle increases, both the decoding accuracy and message extraction rate gradually decrease to the level of random guessing. Due to the infinite visual angle space, attackers are unable to accurately predict a specific visual angle in a three-dimensional scene. Experimentation confirms that even a slight visual angle deviation of 0.1$^\circ$ does not yield reliable results, thus highlighting the concealment efficacy of the proposed scheme in thisletter.

\section{conclusion}
In this letter, we propose a model steganography scheme with implicit neural representation. Using the property of NeRF model to synthesize a new viewpoint image, a secret viewpoint is introduced as a key to get a secret viewpoint image, and then the purpose of delivering secret message is achieved by model steganography. The experimental results prove that this letter adopts any viewpoint in the infinite viewpoint space as the key to ensure the steganography of the scheme, and uses the overfitting training method to realize the backdoor steganography for the message extractor, which ensures the accuracy and the large capacity of the message extracted from the secret viewpoint image.

In this letter, the scheme requires the sender to republish the message extractor every time the message is delivered, which is less secure and more costly than other deep steganography algorithms, and the solution is to equip the message extractor with other image processing functions to enhance the steganographic effect of the network.

\bibliographystyle{IEEEtran}
\bibliography{IEEEabrv,dwn}
\end{document}